%% file: paper.tex
\documentclass[journal]{vgtc}                
\ifpdf
  \pdfoutput=1\relax                   
  \usepackage{graphicx}                
  \DeclareGraphicsExtensions{.pdf,.png,.jpg,.jpeg} 
\else
  \usepackage{graphicx}
  \DeclareGraphicsExtensions{.eps}     
\fi%

\usepackage[caption=false, font=footnotesize]{subfig}

\usepackage{xspace}

\graphicspath{{figures/}{pictures/}{images/}{./}} 

\newcommand{\systemname}{\emph{PipelineProfiler}\xspace}

\usepackage{amsfonts}
\newcommand{\overbar}[1]{\mkern 1.5mu\overline{\mkern-1.5mu#1\mkern-1.5mu}\mkern 1.5mu}

\usepackage{microtype}                 
\PassOptionsToPackage{warn}{textcomp}  
\usepackage{textcomp}                  
\usepackage{mathptmx}                  
\usepackage{amsmath}
\usepackage[shortlabels]{enumitem} 
\usepackage{times}                     
\usepackage{cite}                      
\usepackage{tabu}                      
\usepackage{booktabs}                  
\usepackage{bbm}

\usepackage[ruled,vlined]{algorithm2e} 

\usepackage{xcolor}
\onlineid{1097}
\definecolor{pipelinebluecolor}{RGB}{30,119,180}
\newcommand\cruleblue[3][pipelinebluecolor]{\textcolor{#1}{\rule{#2}{#3}}}
\newcommand\cruleorange[3][orange]{\textcolor{#1}{\rule{#2}{#3}}}

\newcommand\Figure[1]{Fig.~\ref{#1}}

\newcommand\revision[1]{#1}

\newcommand\hide[1]{}
\newcommand\myparagraph[1]{\vspace{4pt}\textbf{#1.}}

\usepackage{tikz}
\newcommand{\critical}[1]{\tikz[baseline=(X.base)]\node [draw=red,fill=red!30,semithick,rectangle,inner sep=2pt, rounded corners=3pt] (X) {Critical};}

\newcommand{\important}[1]{\tikz[baseline=(X.base)]\node [draw=yellow,fill=yellow!30,semithick,rectangle,inner sep=2pt, rounded corners=3pt] (X) {Important};}

\newcommand{\optional}[1]{\tikz[baseline=(X.base)]\node [draw=green,fill=green!30,semithick,rectangle,inner sep=2pt, rounded corners=3pt] (X) {Optional};}

\vgtccategory{Research}
\vgtcpapertype{application/design study}

\title{\systemname: A Visual Analytics Tool for the Exploration of AutoML Pipelines}

\author{Jorge Piazentin Ono, Sonia Castelo, Roque Lopez, Enrico Bertini, Juliana Freire, Claudio Silva}
\authorfooter{All authors are with the New York University. E-mails: \{jorgehpo, s.castelo, rlopez, enrico.bertini, juliana.freire, csilva\}@nyu.edu }

\shortauthortitle{Biv \MakeLowercase{\textit{et al.}}: Global Illumination for Fun and Profit}

\input{text/abstract.tex}


\CCScatlist{ 
 \CCScat{K.6.1}{Management of Computing and Information Systems}%
{Project and People Management}{Life Cycle};
 \CCScat{K.7.m}{The Computing Profession}{Miscellaneous}{Ethics}
}

\teaser{
  \centering
  \includegraphics[width=0.9\linewidth]{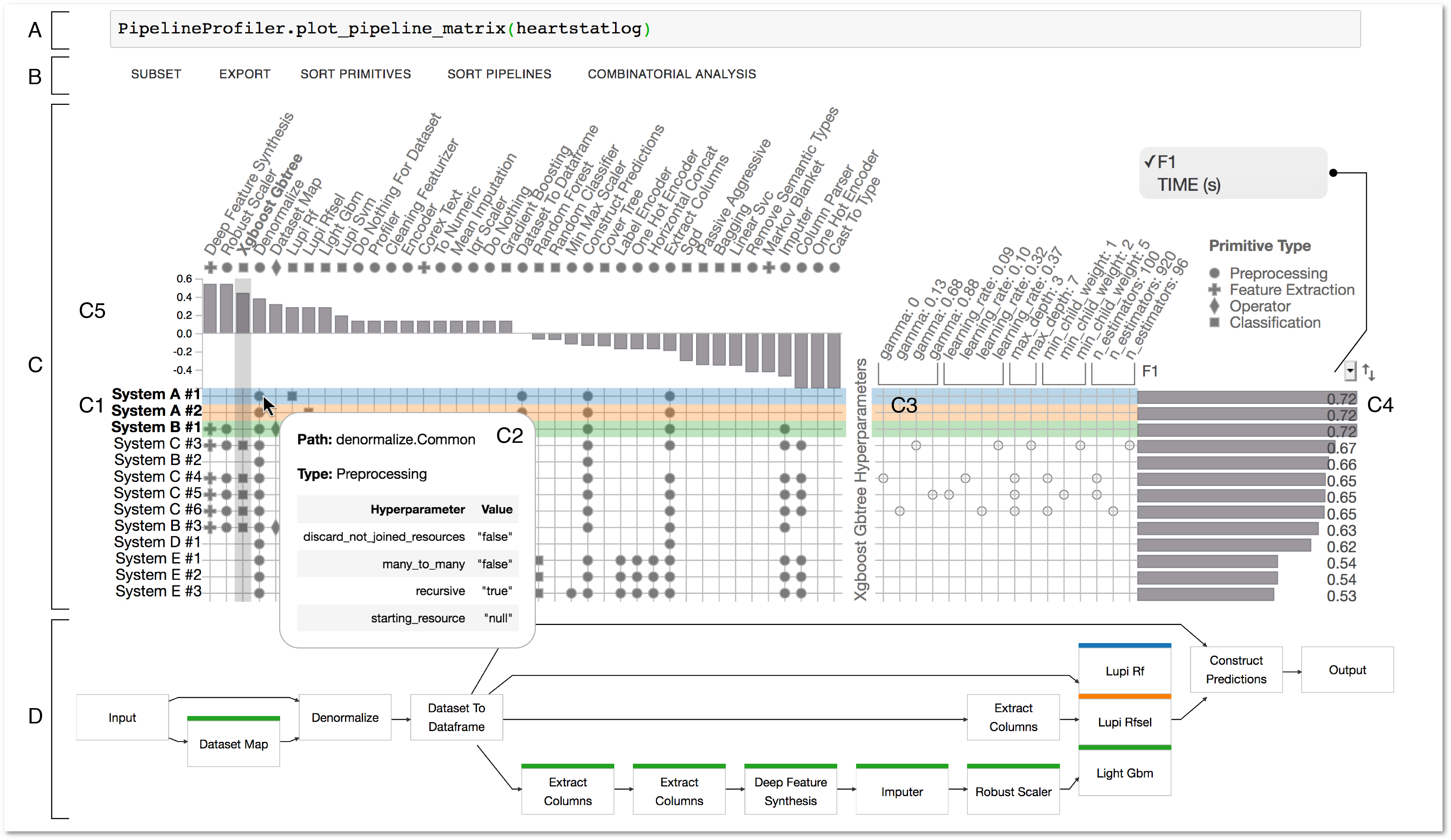}
  \caption{\systemname applied to the analysis of binary classification pipelines generated by five different AutoML systems for the \emph{Statlog (Heart) Data Set}. A) The system is integrated with Jupyter Notebook and can be invoked with one line of code. B) \systemname menu, with options to subset, export, sort, and perform automated analysis on pipelines. C) Pipeline Matrix: C1) Primitives (columns) used by the pipelines (rows).  C2) Tooltip showing the metadata and hyperparameters for a primitive. C3) One-hot-encoded hyperparameters (columns) for the primitive \textit{Xgboost Gbtree} across pipelines (rows).  C4) Pipeline scores: users can select different metrics to rank pipelines. C5) Primitive Contribution View, showing correlations between primitive usage and pipeline scores (here, \textit{Deep Feature Synthesis} has the highest correlation with F1 scores). D) Pipeline Comparison View: visual comparison of the top-3 scoring pipelines.} 
	\label{fig:teaser}
}

\vgtcinsertpkg

\begin{document}



\newcommand\todo[1]{\textcolor{red} { [TODO: {#1}]}}

\input{text/introduction.tex}

\input{text/relatedwork.tex}

\input{text/method.tex}

\input{text/evaluation.tex}

\input{text/conclusion.tex}

\input{text/acknowledgements}

\bibliographystyle{abbrv-doi}

\bibliography{referenceszotero}
\end{document}

%% file: text/abstract.tex
\abstract{
In recent years, a wide variety of automated machine learning (AutoML) methods have been proposed to generate end-to-end ML pipelines. While these techniques facilitate the creation of models, given their black-box nature, the complexity of the underlying algorithms,  and the large number of pipelines they derive, they are difficult for developers to debug. It is also challenging for machine learning experts to select an AutoML system that is well suited for a given problem.
In this paper, we present the \systemname, an interactive visualization tool that allows the exploration and comparison of the solution space of machine learning (ML) pipelines produced by  AutoML systems. \systemname is integrated with Jupyter Notebook and can be combined with common data science tools to enable a rich set of analyses of the ML pipelines, providing users a better understanding of the algorithms that generated them as well as insights into how they can be improved.
We demonstrate the utility of our tool through  use cases where \systemname is used to better understand and improve a real-world AutoML system. Furthermore, we validate our approach by presenting a detailed analysis of a think-aloud experiment with six data scientists who develop and evaluate AutoML tools.
}


%% file: text/introduction.tex
\firstsection{Introduction}
\maketitle

Machine Learning (ML) has been successfully used in a plethora of applications. However, 
assembling end-to-end ML pipelines is a difficult endeavor that requires  a time-consuming, trial-and-error process. This is difficult for ML experts and out of reach for subject-matter experts with little or no training in ML or computer science.
AutoML systems have been proposed to address this challenge.
Given a ML problem, AutoML aims to automate the synthesis of ML pipelines that perform well for the problem  by searching over a space of possible pipelines which use different  combinations of
\revision{primitives -- computational steps / ML algorithms,} and values for their associated hyperparameters \revision{-- settings that configure algorithm behavior} \cite{hutter_automated_2019}. 

AutoML has had substantial practical impact by making data scientists more efficient, enabling researchers to work on harder problems, and democratizing ML to less experienced users~\cite{feurer_auto-sklearn_2019, drori_alphad3m:_2018, feurer2015efficient}. 
Several open-source AutoML systems are currently available.
%
For example, Auto-sklearn \cite{feurer_auto-sklearn_2019} creates pipelines based on a pool of 33 pre-processing and classification primitives. 
In the context of the DARPA Data-Driven Discovery of Models (D3M) program~\cite{elliott_data-driven_2020}, a collaborative effort involving multiple research groups, twenty AutoML systems were developed that synthesize pipelines from a set of over 300 primitives~\cite{d3m_index_primitives_2020,d3m_index_2020}.  
Because exploring the entire space of primitives and associated hyperparameters is not feasible in practice, these systems use sophisticated search strategies to 
to reduce the number of pipelines they need to evaluate. For example, AutoWeka uses Bayesian optimization~\cite{thornton2013auto}, Auto-sklearn uses meta-learning~\cite{feurer_auto-sklearn_2019}, TPOT uses genetic programming \cite{olson_tpot_2019}, and AlphaD3M uses deep learning~\cite{drori_alphad3m:_2018}. 


\myparagraph{Challenges in Understanding and Comparing AutoML Systems} Given the complexity of these systems, two important challenges arise. First, \revision{\emph{it is difficult to evaluate the efficiency, correctness and performance or AutoML systems}}.  
To debug  an AutoML system, developers must analyze logs consisting of synthesized pipeline instances and their outcomes. They need to assess, for instance, the efficiency of the search process, the structural diversity of the derived pipelines, how well  the search covers the available primitives, and whether primitives were used correctly.
This requires involved analyses that are further complicated due to the fact that not only do these logs contain a large number of instances, but the instances can have a complex structure and use a wide variety of primitives.
%
%

The second challenge emerges from the \emph{need to compare different AutoML systems}. Given the growing number of available systems~\cite{google_cloud_2020, mljar_machine_2020, ibm_watson_2020, drori_alphad3m:_2018, olson_tpot_2019, feurer_auto-sklearn_2019}, to make an informed decision when selecting a system,  it is important for users to understand how well the systems perform for different problems, and identify features that contribute to a system being more or less effective than others for a task. Insights obtained in such a comparison are also of great value for AutoML developers, as these may help them improve their systems. 

Visual analytics techniques have been proposed to make AutoML more transparent 
by enabling the exploration of the produced pipelines, the primitives employed, and their associated hyperparameters~\cite{wang_atmseer:_2019, golovin_google_2017, park_visualhypertuner_2019, weidele_autoaiviz_2020, cashman_ablate_2019}. 
%
However, these have important limitations with respect to the challenges we outlined: (1) the graphical encodings used for pipelines only support fixed templates, thus they cannot explore pipelines that have complex structure; and (2) they do not support the comparison of multiple AutoML systems.
%
%
For example: AutoAIViz~\cite{weidele_autoaiviz_2020} only supports sequential pipelines that consist of  three primitives: two transformers (e.g., pre-processing or feature selection) and one estimator (e.g., classification or regression); ATMSeer \cite{wang_atmseer:_2019} presents information only for the estimator step of the pipeline. 
%
%
Using these approaches, it is not possible to explore pipelines that have a complex structure such as directed acyclic graphs (DAG), or long pipelines that contain multiple estimators, such as the ones used for  ensemble models which have been shown to outperform simpler pipelines~\cite{feurer_auto-sklearn_2019}. 
%
%

It is worth noting that since these techniques are  tightly coupled with a specific AutoML system,  they do not address the challenges involved in comparing different systems. 
For example, while ATMSeer \cite{wang_atmseer:_2019} visualizes pipelines produced by ATM~\cite{swearingen_ATM_2017}, AutoAIViz can only show pipelines from AutoAI~\cite{wang2019human}. 


\hide{
\begin{figure}[tb]
    \centering
    \includegraphics[width=\columnwidth]{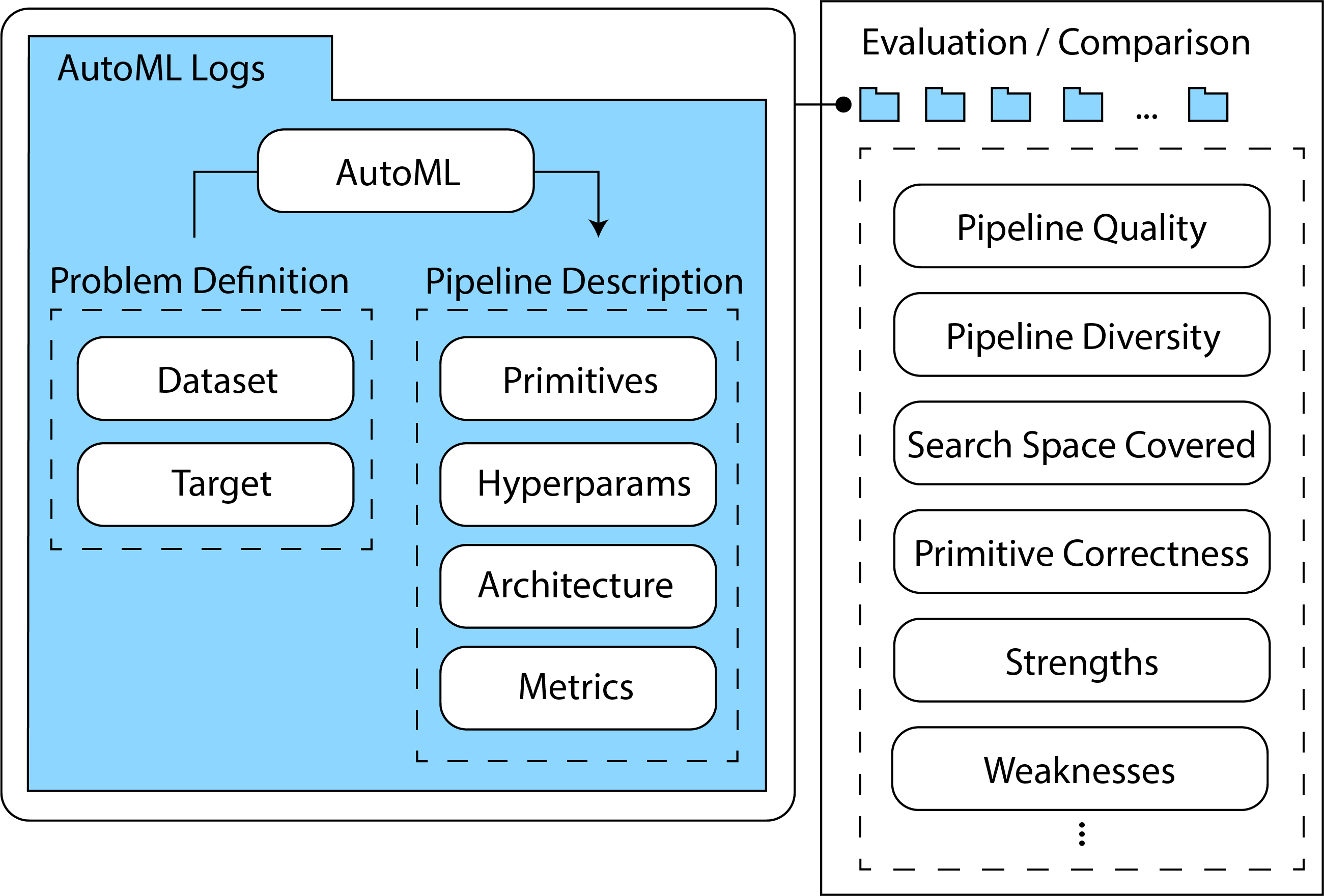}
    \caption{Overview of AutoML systems investigation. Left: AutoML logs, consisting of dataset metadata and pipeline output description. Right: Key factors for AutoML Evaluation and Comparison.}
    \label{fig:devworkflow}
\end{figure}{}
}

\myparagraph{Our Approach} We propose \systemname, a visual analytics tool that enables the exploration and comparison of end-to-end ML pipelines produced by multiple AutoML systems. \revision{\systemname takes as input pipelines represented with a common description language, consisting of pipeline architecture -- the set of primitives and data flow between primitives (encoded as a Directed Acyclic Graph); and pipeline metadata -- hyperparameters, running time, evaluation scores, etc.} \Figure{fig:teaser} shows the main components of the system.
The Pipeline Matrix provides a visual summary of a collection of AutoML pipeline instances that captures structural information, including the primitives and associated hyperparameters, used in the pipelines, as well as the outcome of the pipeline encoded in a score.
This representation is compact and can effectively encode pipelines derived by multiple AutoML systems that have both complex structure and use a variety of primitives. By showing the correlations between primitive usage and pipeline scores and how much primitives contribute to the score, the representation can also help uncover insights into the suitability (or effectiveness) of  primitives for specific data and problem types.
Users can drill down into the structural details of pipelines, and examine both their differences and similarities. As we discuss later, these analyses enable the identification of patterns which can, for example, expose interesting aspects of the search strategies used by AutoML systems.
\systemname is integrated with Jupyter Notebooks. This enables complex analyses to be performed
over pipeline collections that leverage the rich ecosystem of Python tools for data science. 


Our design was inspired by requirements from developers of AutoML systems and experts that evaluate and compare these systems in the context of the DARPA D3M project~\cite{elliott_data-driven_2020}. 
These experts used \systemname and provided feedback throughout its development.

Our main contributions can be summarized as follows:
\begin{itemize}     \vspace{-.15cm}
    \item We propose \systemname, which is, to the best of our knowledge the first visual analytics tool for exploring and comparing  pipelines produced by multiple AutoML systems. 
    \vspace{-.15cm}
    \item \revision{The tool combines effective visual representations and interactions for AutoML analysis}, including: the Pipeline Matrix, a visual encoding that summarizes the pipelines, primitives and hyperparameter values produced by  AutoML systems; and the Graph Comparison View, which applies graph matching to highlight the structural differences and similarities among a set of pipelines of interest. 
        \vspace{-.15cm}
    \item To derive the information needed to convey the correlation between primitives and pipeline scores, we propose 
    a new algorithm that identifies groups of primitives that have a strong impact on pipeline scores. 
        \vspace{-.15cm}
    \item To demonstrate the effectiveness of \systemname, we present two use cases that show how the tool was used to obtains insights that resulted in efficiency improvements to an open-source AutoML system.
    We  also discuss a detailed analysis of a think-aloud experiment with six data scientists, which suggests that the tool is both usable and useful, and highlights the benefits of the integration of the tool with Jupyter Notebooks both in the flexibility it affords for complex analyses and the simplicity in incorporating the tool into the experts' development and evaluation workflows.
\end{itemize}


%% file: text/relatedwork.tex
\section{Related Work} \label{sec:rw}


AutoML has emerged as an approach to simplify the use of ML for different applications, and many systems that support AutoML are currently available~\cite{hutter_automated_2019, google_cloud_2020, mljar_machine_2020, ibm_watson_2020, drori_alphad3m:_2018, olson_tpot_2019, feurer_auto-sklearn_2019}. 
While early work on AutoML focused on hyperparameter optimization and on ML primitives, recent approaches aim to efficiently automate the synthesis of end-to-end pipelines -- from data loading, pre-processing, feature extraction, feature selection, model fitting and selection, and hyper-parameter tuning~\cite{swearingen_ATM_2017, drori_alphad3m:_2018, shang_democratizing_ml_2019, feurer_auto-sklearn_2019, olson_tpot_2019}. 
AlphaD3M uses deep learning to learn how to incrementally construct ML pipelines, framing the problem of pipeline synthesis for model discovery as a single-player game with a neural network sequence model and Monte Carlo Tree Search (MCTS)~\cite{drori_alphad3m:_2018}. 
Auto-Sklearn produces pipelines using Bayesian optimization combined with meta-learning~\cite{feurer_auto-sklearn_2019}, and TPOT uses genetic programming and tree-based optimization~\cite{olson_tpot_2019}. 


%
During the search for well-performing pipelines, AutoML systems can generate a large number of  pipelines. This makes the analysis of the search results a challenging problem, in particular when pipelines have similar scores. As a consequence, the selection of the best performing end-to-end pipeline becomes expensive, time-consuming and a tedious process. In the following, we discuss research that has attempted to tackle this challenges through visualization.

\myparagraph{Explaining AutoML}
The black box nature of AutoML systems and the difficulty in understanding the inner-workings of these systems lead to reduced trust in the pipelines they produce~\cite{wang_atmseer:_2019}. There have been some attempts to make the AutoML process more transparent through insightful visualizations of the resulting pipelines. These can be roughly grouped into two categories: hyperparameter visualization~\cite{wang_visual_2019, park_visualhypertuner_2019, golovin_google_2017} and pipeline visualization \cite{weidele_autoaiviz_2020,cashman_ablate_2019}. 


ATMSeer~\cite{wang_atmseer:_2019} and Google Vizier~\cite{golovin_google_2017} are approaches to visualize hyperparameters. ATMSeer, which is integrated with the ATM AutoML framework~\cite{swearingen_ATM_2017}, displays the predictive model (i.e., last step of the ML pipeline) together with its hyperparameters and performance metrics to the user. Users can use  crossfiltering  (e.g., over ML algorithms) to facilitate the exploration of large collection of pipelines and refine the search space of the AutoML system, if needed. 
%
Google Vizier 
makes use of a parallel coordinate view where hyperparameters and objective functions are displayed to the users. It allows them to examine how the different hyperparameters (dimensions) co-vary with each other and also against the objective function.
%
Although these methods help AutoML users to analyze the generated pipelines, most of them only support the analysis of its last step which is the fitted model, leaving aside important aspects of the pipeline such as data cleaning and feature engineering. In contrast, \systemname provides a visualization to explore and analyze the end-to-end pipeline -- from data ingestion and engineering to model generation. 


Systems that support pipeline visualization include  AutoAIViz~\cite{weidele_autoaiviz_2020} and REMAP~\cite{cashman_ablate_2019}. AutoAIViz uses Conditional Parallel Coordinates (PCP)~\cite{weidele_conditional_2019} to represent sequential pipelines and their hyperparameters. The system provides a hierarchical visualization which shows the pipeline steps (on the first level of PCP) as well as the hyperparameters of each step (on the second level of PCP). REMAP~\cite{cashman_ablate_2019} focuses on pipelines that use deep neural networks. It proposes a new glyph, called Sequential Neural Architecture Chips (SNAC), which shows both the layer type and dimensionality, and allows users to interactively add or remove layers from the networks. Despite their ability to show end-to-end pipelines, both systems can only show linear pipelines, making it difficult to explore pipelines created by different AutoML systems that have more complex structure. 
Furthermore, REMAP was designed specifically to visualize neural network architectures, and thus it is not suitable to explore general ML pipelines that use different learning techniques.
In this work, our goal is to allow users to explore, compare and analyze pipelines generated by multiple AutoML systems which can have nonlinear pipeline structures and use a variety of primitives and learning techniques.

\myparagraph{Visual Analytics for Model Selection}
%
Selecting a \emph{good} model among the potentially large set of models (or pipelines) derived by an AutoML system is a challenging problem that has attracted significant attention in the literature.
%
Visual analytics systems such Visus~\cite{santos_visus_2019}, TwoRavens~\cite{gil_tworavens_2019}, and Snowcat \cite{cashman_snowcat_2018} provide a front-end to AutoML systems and guide subject-matter experts without training in data science  through the model selection process. They focus on providing explanations for models, and some provide
simple mechanisms to compare models (e.g., based on the scores, or the actual explanations).
%
Other approaches focus exclusively on model selection.
RegressionExplorer~\cite{dingen_regressionexplorer:_2018} enables the creation, evaluation and  comparison of logistic regression models using subgroup analysis. 
%
%
ModelTracker \cite{amershi_modeltracker:_2015} and Squares~\cite{ren_squares:_2017} introduce novel encodings to investigate different models, by visually comparing histograms based on the statistical performance metrics of binary and multi-class classifiers. They also show instance-level distribution information and enable multi-scale analysis.  \revision{For a more complete list of visual analytics tools for interactive model exploration, we refer the interested readers to the following surveys  \cite{liu2017towards, garcia2018task, hohman2018visual, spinner2019explainer}}.
The majority of these methods were designed to evaluate and select predictive models based on  the performance results. However, they do not take into account additional metrics like  \emph{running time}, i.e., how long pipelines take to run, or \emph{primitive usage}, i.e., whether primitives are used correctly and effectively.
\systemname not only encodes this information in a compact visual
representation, but it also provides a usable interface that allows users interact with a pipeline collection at different levels of abstraction -- from a high-level overview to drilling down to inspect details of select pipelines. 

\myparagraph{Interactive Model Steering}
%
Systems such as TreePOD \cite{muhlbacher_treepod_2018}, BEAMES \cite{das_beames_2019} and EnsembleMatrix \cite{talbot_ensemblematrix:_2009} support the analysis and refinement of models through interactive visualizations, and allow users to explore the effects of modifying some parameters. BEAMES~\cite{das_beames_2019} lets users steer the training of new regression models from a set of previous models. 
It presents model performance information to the users and trains new models based on the user feedback. 
EnsembleMatrix \cite{talbot_ensemblematrix:_2009} and TreePOD~\cite{muhlbacher_treepod_2018} enable users to steer the creation of ensemble decision tree models. With EnsembleMatrix, users can combine and choose weights for decision trees and interactively evaluate the performance of the ensemble model. Conversely, TreePOD supports the optimization of multiple objectives, including performance and interpretability. 
Although these systems help users understand the impact of these parameters over the models, they do not consider other relevant steps (also called primitives) that are part of end-to-end pipelines like data ingestion and feature engineering, which could have a significant impact in the final model performance.  The primitives contribution view in \systemname displays the correlations between
primitive usage and scores, allowing users to infer which primitives can lead to well-performing pipelines. Users can then drill down and further explore individual pipelines, their primitives and hyperparameters.

%% file: text/method.tex
\section{\systemname: exploring end-to-end ML pipelines} \label{sec:method}

In this section, we describe \systemname, a tool that enables the exploration of end-to-end machine learning pipelines produced by AutoML systems. We first present the desiderata we distilled from  interviews with AutoML experts and subsequently used to guide our design choices. Then, we describe the components of \systemname, how they are integrated, and the algorithms we developed to enable the effective analysis of ML pipelines. Finally, we briefly describe the implementation details of our system.


\subsection{Domain Requirements}
\label{sec:requirements}


We conducted interviews with six data scientists who actively work with AutoML systems in the context of the D3M project~\cite{elliott_data-driven_2020}: the developers of four distinct AutoML systems (D1 - D4) and two data scientists that are tasked with evaluating the D3M AutoML systems (E1 and E2). Since each developer works on a specific system, they have different needs  and follow distinct workflows. However, they share some challenges. The AutoML evaluators are part of the D3M management team. They are  responsible for selecting what types of ML tasks the developers must focus on and also evaluate system performance. 

D3M pipelines are represented as JSON-serialized objects that contain metadata, input and output information, and the pipeline architecture,  which is described as a directed acyclic graph (DAG)~\cite{milutinovic_2019,d3m_datadrivendiscovery_2020}. 
The exploration of  ML pipelines collections is a task performed by all AutoML developers and evaluators. All interviewees said they explored pipelines by looking at their JSON representations, and complained that reading the text files, and inspecting the pipelines
one at a time was a tedious and time-consuming task. 
Understanding and comparing the pipelines is difficult, in particular, since the DAG structure is hard to grasp from the JSON representaion. 

D1 said she does not have time to inspect pipelines often, and instead focuses on assessing cross-validation scores and looking for correlations between primitives and performance scores. In contrast, D2, D3 and D4 they examine the pipeline DAGs and their architecture. D1 and D2 also analyze the prediction and training time. They mentioned that their AutoML systems were evaluated within a given time budget, therefore training time is an important metric for them. 

D2's system has a blacklisting feature: when a primitive is found to have poor performance, it can be flagged and excluded from the search process.
Therefore, he was interested in identifying when a primitive was associated with high or low pipeline scores. 

D3 usually compares pipelines using their cross-validation scores. When he finds a problem for which his systems derives sub-optimal pipelines, he inspects the pipelines derived by other systems. His goal is to understand which features in his pipelines lead to the low scores, and conversely, why the pipelines derived by the other systems perform better. By answering these questions, he hopes to gain insights into if and how he can improve his system. 
He is also interested in exploring the pipelines at the  hyperparameter level, but said this is currently not possible due to the large number of primitives (over 300), pipelines, and parameters involved. 

D4 is also interested in comparing pipelines, albeit for a different reason. More specifically, he is interested in comparing AutoML pipelines from different sources, including human generated pipelines. His goal is to evaluate if there are differences between machine and human-generated pipelines. He is also interested in primitive similarity. More specifically, he wants to find which primitives are exchangeable within a pipeline architecture.

The analysis workflow followed by the AutoML evaluators is significantly different from that of the developers. While developers focus on pipeline structure, evaluators are mostly concerned with how well the systems perform and the problem types (e.g., classification, regression, forecasting, object detection, etc.) they currently support and should support  in future iterations. More specifically, E1 and E2 said that their workflow consisted mostly on evaluating AutoML systems based on their cross-validation scores. However, they were also interested in checking how the primitives were being used, and whether AutoML systems produced 
different pipelines a given problem type. More specifically, they stated that if all AutoML systems derived the same (or very similar) pipelines, the task they are solving is no longer challenging and new problem types should be proposed. For formal evaluations, the D3M systems are evaluated using sequestered problems that are not visible to the developers. Thus,  to give actionable insight to AutoML developers without disclosing specifics of the sequestered problems, E2 was also interested in identifying why pipelines fail.

We compiled the following desiderata from the interviews. \revision{Each requirement is marked as Critical, Important, or Optional, according to the number of users that requested it, and how they guided the design of our tool.}

\begin{enumerate}[start=1,label={[\bfseries R\arabic*]}]
\item \label{req:summary} Pipeline collection overview and summary: all participants would like to visualize and compare multiple pipelines simultaneously, instead of inspecting them one by one \critical. 
\vspace{-.15cm}
\item \label{req:primitive_usage} Primitive usage: E1 and E2 are interested in exploring how primitives are used across different AutoML systems. More specifically, they want to check if the systems are generating diverse solutions and if there are underutilized primitives \important.
\vspace{-.15cm}
\item \label{req:hyperparameters} Visualizing primitive hyperparameters: D3 would like to be able to explore the hyperparameter space of the primitives used in his pipelines \optional.
\vspace{-.15cm}
\item \label{req:metadata} Visualizing pipeline metadata: D1, D2, E1 and E2 mentioned they were interested in visualizing and comparing different aspects of the trained pipelines, including scores, prediction and training time \important. 
\vspace{-.15cm}


\item \label{req:correlations} Finding correlations between primitives and scores: D1 and D2 were interested in identifying primitives that correlate with high scores on different problems and datasets. Furthermore, D2 would like to see primitives that perform poorly in order to blacklist them, and E2 is interested in identifying possible causes for pipeline failure (i.e., low scores) \important.
\vspace{-.15cm}

\item \label{req:graphdiff} Visualizing and comparing pipeline graphs: all developers were interested in visualizing the connections between pipelines primitives using a graph metaphor. Furthermore, D3 and D4 are interested in performing a detailed comparison of the pipeline graphs. In particular, they want to identify how different AutoML systems structure their pipelines to solve a particular problem type \critical. 
\vspace{-.15cm}
\end{enumerate}

\subsection{Visualization Design}


In order to fulfill the requirements identified in the previous section, we developed \systemname, a tool that enables the interactive exploration of pipelines generated by AutoML systems. 
\Figure{fig:teaser} shows \systemname being applied to compare pipelines derived by five distinct AutoML systems for a classification problem that aims to predict the presence of heart disease using the \emph{Statlog (Heart) Data Set} \cite{dua2017uci}. 
%
The main components of \systemname are the Pipeline Matrix (C) and the Pipeline Comparison View (D).  
The Pipeline Matrix (C) shows a tabular summary of all the pipelines in the collection. The user can also drill down and explore one or multiple pipelines in more detail -- the graph structure of selected pipelines are displayed in the Pipeline Comparison View (D) upon request. 
The system Menu (B) enables users to focus on a subset of the pipelines, export pipelines of interest to Python, sort the table rows and columns, and perform automated analyses over groups of primitives. These operations are described later in this section. \systemname is implemented as a Python library that can be used with Jupyter Notebooks to facilitate the integration with the workflow of the AutoML community (A). 

%


\subsubsection*{Pipeline Matrix}

The Pipeline Matrix provides a summary for a collection of machine learning pipelines~\ref{req:summary} selected by the user.
Its visual encoding was inspired by visualizations used for topic modeling systems~\cite{chuang2012termite, alexander2014serendip}. However, instead of words and documents, this matrix represents whether a primitive is used \ref{req:primitive_usage} in a  machine learning pipeline (\Figure{fig:teaser}(C1)). 
Users can interactively reorder rows and columns  according to pipeline evaluation score, pipeline source (AutoML system that generated it), primitive type (e.g., classification, regression, feature extraction, etc.), and estimated primitive contribution (i.e., correlation of primitive usage with pipeline scores). Furthermore, we use shape to encode primitive types. \revision{When columns are sorted by primitive type, we include vertical separator  lines between primitive types to help differentiate groups of primitives. A similar design was used in \cite{telea2006combining} to highlight groups of samples.} 

To support the exploration of hyperparameters~\ref{req:hyperparameters}, \systemname  implements two interactions that show this information on demand: parameter tooltip and one-hot-encoded parameter matrix. When the user hovers over a cell in the matrix, a tooltip shows the primitive metadata (type and Python path) as well as a table with all the hyperparameters set. \Figure{fig:teaser}(C2) shows a tooltip for primitive \emph{Denormalize}, with four hyperparameter values set. 
Users can also inspect a summary of  hyperparameter space for a primitive by selecting a column in the Pipeline Matrix. When a primitive (column) is selected, all of its hyperparameters are represented using a one-hot-encoding approach: each hyperparameter value becomes a column in the matrix, and dots indicate when the hyperparameter is set in a pipeline. \revision{Since ML researchers are familiar with one-hot-encoding, this representation is natural for them. } \Figure{fig:teaser}(C3) shows the hyperparameter space of  \emph{Xgboost Gbtree}.

Domain experts were interested in exploring pipeline metadata~\ref{req:metadata}, including  training and testing scores, training time and execution time. \systemname shows the pipeline metadata in the Metric View (\Figure{fig:teaser}(C4)). Users can select which metric to display using a drop down menu, and the numerical values are shown in bar chart aligned with the matrix rows. In C4, the user can choose to display the metric \emph{F1} or the prediction time. Pipeline rows can be re-ordered based on the metric, and to enable a comparison across systems, users can also interactively group pipelines based by the system that generated them.


To convey information about the relationships between primitive usage and pipeline scores~\ref{req:correlations}, we designed the Primitive Contribution view.
This view shows an estimate of how much a primitive contributes to the score of the pipeline using a bar chart encoding,  aligned with the columns of the matrix (\Figure{fig:teaser}(C5)). The contribution can be either positive or negative, representing positive or negative primitive correlation with the scores. For example, in C5, \emph{Deep Feature Synthesis} is the primitive most highly correlated with \emph{F1}.

We estimate the primitive contribution using the Pearson Correlation (PC) between the primitive indicator vector $p$ ($p_i=1$ if pipeline $i$ contains the primitive in question and $p_i=0$ otherwise)  and the pipeline metric vector $m$, where $m_i$ is the metric score for pipeline $i$. Since $p$ is dichotomous and $m$ is quantitative, PC can be computed more efficiently with the Point-Biserial Correlation (PBC) coefficient~\cite{sheskin2003handbook}. PBC is equivalent to the Pearson correlation, but can be evaluated with fewer operations.  
%
Let $\overbar{m_1}$ be the mean of the metric score ($m$) when the primitive is used ($p_i=1$);  $\overbar{m_0}$, the mean of the scores when the primitive is not used ($p_i=0$); $s$ be the standard deviation of all the scores ($m$); $n_1$ be the number of pipelines  where the primitive is used; $n_0$ be the number of pipelines where the primitive is not used; and $n=n_1+n_2$. The Point-Biserial Correlation is computed as: $
PBC = \left[\frac{\overbar{m_1} - \overbar{m_0}}{s}\right] \sqrt{\frac{n_1 n_0}{n^2}}
$

\revision{The choice of Pearson Correlation (PC) as a proxy for primitive contribution is motivated by its natural interpretation, community familiarity, and fast computation. However, PC is sensitive to sample size and outliers~\cite{kim2015instability}.
There are more robust methods to evaluate the importance of hyperparameters/attributes \cite{hutter2014efficient,van2018hyperparameter, bailis2017macrobase}, but they are not tailored for interactive systems. In particular, these algorithms have longer running times and require data pre-processing, i.e., 
training regression models \cite{hutter2014efficient, van2018hyperparameter} or computing frequent patterns~\cite{bailis2017macrobase}.}

\subsubsection*{Pipeline Comparison View}

To provide a concise summary of a collection of pipelines, the Pipeline Matrix models the pipelines as a set of primitives that can be effectively displayed in a matrix.
However, while analyzing pipelines collections, AutoML developers also need to examine and compare
the graph structure of the pipelines~\ref{req:graphdiff}. 
The Pipeline Comparison view (\Figure{fig:teaser}(D)) consists of a node-link diagram that shows either an individual pipeline, or visual-difference summary of multiple pipelines selected in the matrix representation. In the summary graph,  each primitive (node) is color-coded to indicate the pipeline where it appears. If a primitive is present in multiple pipelines, all corresponding colors are displayed. If a primitive appears in all selected pipelines, no color is displayed. 

\begin{figure}[tb]
     \centering
     \subfloat[Best performing pipeline \cruleblue{0.2cm}{0.2cm} (F1 Macro: 0.45) \label{fig:20news_best}]{
         \centering
         \includegraphics[width=\linewidth, height=1.4cm]{pictures/20news_top_v3.png}
     }
     \qquad
     \subfloat[Worst performing pipeline \cruleorange{0.2cm}{0.2cm} (F1 Macro: 0.06)
     \label{fig:20news_worst}]{
         \centering
         \includegraphics[width=7.8cm, height=1.4cm]{pictures/20news_bottom_v3.png}
     }
     \qquad
     \subfloat[Merged pipeline \label{fig:20news_merged}]{
         \centering
         \includegraphics[width=\linewidth]{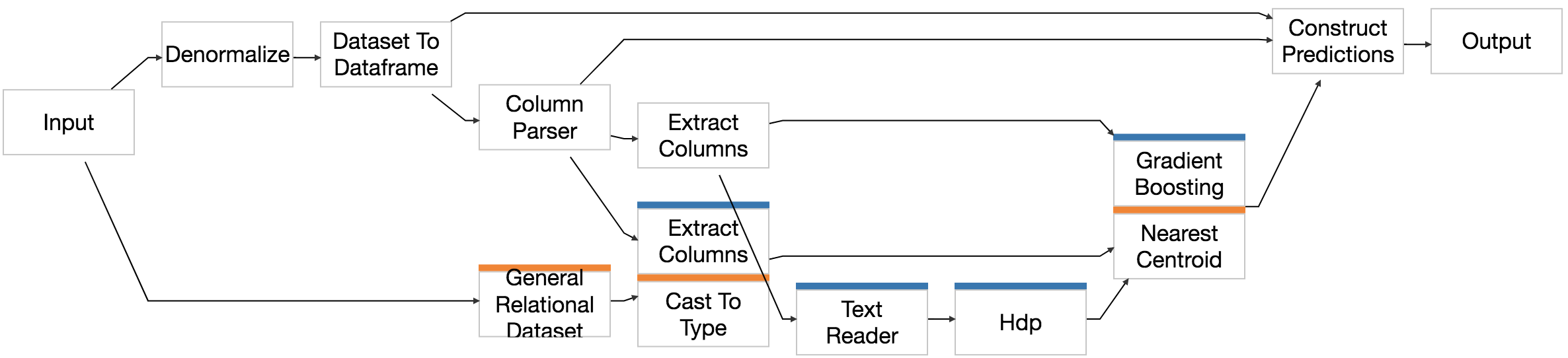}
     }
     \vspace{-.15cm}
        \caption{Pipeline Comparison View, showing the best and worst pipelines for a multitask classification problem on the  \emph{20 newsgroups} dataset. (a) and (b) show individual pipeline structure for the best and worst pipelines respectively. (c) presents the merged view of both pipelines, highlighting the differences between them using color-coded headers. 
        }
       \label{fig:pipelineComparisonView}
     \vspace{-.45cm}
\end{figure}

The Pipeline Comparison View enables users to quickly identify similarities and differences across  pipelines. \Figure{fig:pipelineComparisonView} shows the best (a) and worst pipelines (b) solving the \emph{20 newsgroups classification problem}~\cite{lang1995newsweeder},  
and a merged pipeline (c) that highlights the differences between the two pipelines, clearly showing that
the best pipeline (blue) uses  a Gradient Boosting classifier, and an HDP and Text Reader feature extractors.

To support the comparison of multiple pipeline structures, we adapted the graph summarization method proposed by Koop et al.~\cite{koop_visual_2013}. 
Since ML pipelines are directed-acyclic graphs, we modify the method to avoid cycles in the merged graph. The algorithm creates a summary graph by iteratively merging graph pairs. The merge of two graphs $G_1$ and $G_2$ is performed in four steps, as detailed below.



\emph{1) Computing Node Similarity:} Let $p \in G_1$ and $q \in G_2$ be two primitives (nodes). We say that $p$ and $q$ have the same type if they perform the same data manipulation (e.g., Classification, Regression, Feature Extraction, etc.). The similarity $f$ is given by:

\vspace*{-15pt}
$$
f(p, q) = \begin{cases}
1.0 & \text{if $p$ and $q$ are the same primitive (total match)}\\
0.5 & \text{if $p$ and $q$ have the same type (partial match)}\\
0.0 & \text{otherwise}
\end{cases}
$$
\vspace*{-10pt}

As in Koop et al. \cite{koop_visual_2013}, we use the Similarity Flooding algorithm~\cite{melnik2002similarity} to iteratively adjust the similarity between nodes and take node connectivity into account. We refer the reader to \cite{melnik2002similarity} for details.

\emph{2) Graph Edit Matrix Construction: } In order to match two graphs, $G1$ and $G2$, the algorithm builds a graph edit matrix that contains the all the possible costs to transform $G1$ into $G2$. Let $m$ and $n$ be the number of nodes $G_1$ and $G_2$ respectively. The edit matrix $E$ is defined so that the selection of one entry from every row and one entry from every column corresponds to a graph edit that transforms $G_1$ into $G_2$ \cite{riesen2009approximate}. $E$ contains the costs to add ($a$), delete ($d$) and substitute ($s$) nodes. We choose costs that prioritize node substitutions in case of a total or partial match: $s_{i,j} = 1-f(i,j)$, $d = 0.4$ and $a=0.4$.

\renewcommand{\arraystretch}{0.6}
{
\footnotesize
$$
E = \left[
\begin{array}{cccc|cccc}
s_{00} & s_{01} & \ldots & s_{0n} & a_0 & \infty & \ldots & \infty \\
s_{10} & s_{11} & \ldots & s_{1n} & \infty & a_1 & \ldots & \infty \\
\vdots & \vdots & \ddots & \vdots & \vdots & \vdots & \ddots & \vdots \\
s_{m0} & s_{m1} & \ldots & s_{mn} & \infty & \infty & \ldots & a_m \\
\hline
d_0 & \infty & \ldots & \infty & 0 & 0 & 0 & 0 \\
\infty & d_1 & \ldots & \infty & 0 & 0 & 0 & 0 \\
\vdots & \vdots & \ddots & \vdots & \vdots & \vdots & \ddots & \vdots \\
\infty & \infty & \ldots & d_n & 0 & 0 & 0 & 0 \\
\end{array} 
\right]
$$
}

\emph{3) Node matching: } We use the Hungarian algorithm~\cite{kuhn1955hungarian} to select one entry of every row and one entry of every column of $E$, while minimizing the total cost of the graph edit. Two nodes match when one can be substituted by the other, i.e., their substitution entry is selected from the matrix.

\emph{4) Graph merging: } We merge $G_1$ and $G_2$ by creating a compound node for every pair of nodes that were matched in step 3. However, since machine learning pipelines are directed acyclic graphs, we do not want the merged graph to have cycles either. Therefore, we use the additional constraint to only merge nodes that do not result in cycles in the merged graph. This check is done using a depth search first after each merge.



\subsubsection*{Combined-Primitive Contribution}

The primitive contribution  presented in the previous section does not take into account primitive interactions. For example, it might be the case that for a given problem, the classification algorithm \emph{SVM} and the pre-processing \emph{PCA} together produce good models , but they may lead to low-scoring pipelines when used independently. Because the contribution is estimated with the Point-Biserial Correlation of the binary primitive usage vector and pipeline score, interactions involving multiple primitives are not considered.

To take all primitive interactions into consideration, it would be necessary to check for the correlations of all the primitive groups in the powerset of our primitive space. This strategy has two critical problems: 1) it is not computationally tractable, and 2) it would result in a number of combinations prohibitively large for users to inspect. To tackle this challenge, we propose a new algorithm to identify groups of primitives strongly correlated with pipeline scores. The algorithm works as follows: for every combination of primitives $S$ up to a predefined constant size, 1) create a new primitive indicator vector $\hat{p} = \prod\limits_{i \in S}p_i$, which contains 1 if the set of primitives is used in the pipeline, and 0 otherwise. 2) compute the correlation of the primitive group with the pipeline scores using the Point-Biserial Correlation. 
3) select which combination of primitives to report to the user. We only report the primitive group if its Pearson correlation is greater than the Pearson correlation of all the elements in its powerset. Algorithm \ref{alg:cpc} describes CPC in detail.

\begin{algorithm}
 \KwIn{$p_1, p_2, \ldots, p_n$, the primitive indicator vectors \newline 
 $m$, the evaluation metric score vector \newline
 $K$, the maximal cardinality of the primitive group}
 $I \gets [1, 2, \ldots, n]$\; 
 $contributions \gets \{\}$\;
 \tcp{Computing group correlations}
 \tcp{$2_{\leq K}^I$ : powerset of I up to cardinality K}
 \For{$S \in 2_{\leq K}^{I}$} {
 $\hat{p} \gets \prod\limits_{i \in S}p_i$\;
 $contributions[S] \gets corr(\hat{p}, m)$\;
 }
 \tcp{Selecting primitive groups to report (R)}
 $R \gets [$ $]$\;
 \tcp{$2_{\geq 2, \leq K}^{I}$ powerset of I of cardinality $c,  2\leq c \leq K$}
 \For{$S \in 2_{\geq 2, \leq K}^{I}$}{
   $keep \gets $ True\;
   \tcp{Checks if there is a subset of S with greater contribution}
   \For{$sub \in 2_{\geq 1}^{S}$}{
     $a \gets contributions[S]$\;
     $b \gets contributions[sub]$\;
     \If{$|b| \geq |a|$}{
       $keep \gets $ False\;
       break\;
     }
   }
   \If{$keep = $ True}{
     $R \gets R \cup S$\;
   }
 }
 \Return R
\caption{Combined-Primitive Contribution}
\label{alg:cpc}
\end{algorithm}

\begin{figure}[tb]
    \centering
    \includegraphics[width=0.9\columnwidth]{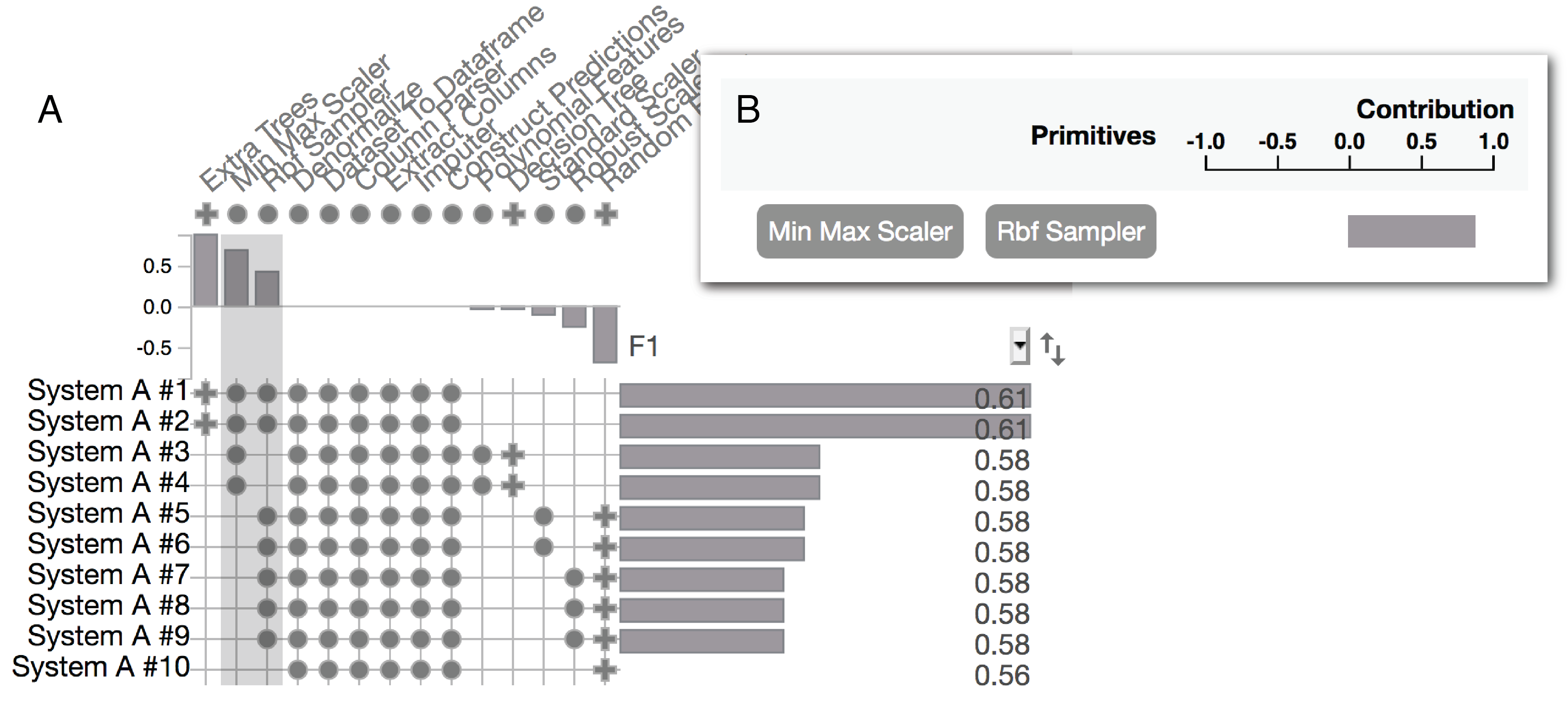}
    \vspace{-.45cm}
    \caption{Combined-Primitive Contribution applied to pipelines that solve a classification problem on the Diabetes \cite{dua2017uci} dataset: A) The Pipeline Matrix representation of the pipeline collection. B) The Combinatorial Analysis View, showing a group of two primitives, \emph{Min Max Scaler} and \emph{RBF Sampler}, that correlate with higher F1 scores. The two primitives are highlighted in (A). Notice that pipelines with higher scores use both primitives (\#1, \#2) -- pipelines that use them separately have lower scores (\#3 - \#9).}
    \label{fig:combinatorial_analysis_view_example}
\end{figure}

The idea behind CPC is simple. The algorithm checks the correlation between combinations of primitives and the pipeline scores, and reports surprising combinations to the user (correlations not shown in the Primitive Contribution View). The user defines $K$ (in our tests, we found that $K=3$ is effective \revision{; i.e. larger values did not return more patterns in the data}). If there are $n$ primitives, the algorithm evaluates $\sum\limits_{k=1}^{K}{n \choose k}$ groups of primitives and has a time complexity of $O\left({n \choose K}\right)$. In \systemname, this CPC can be run via the ``Combinatorial Analysis'' menu (\Figure{fig:teaser}(A)). When the algorithm is run, we show a table containing the selected groups of primitives and the correlation values.  \Figure{fig:combinatorial_analysis_view_example} shows an example of a CPC run over pipelines derived to perform a classification task using the Diabetes dataset~\cite{dua2017uci}. \revision{When a primitive group is clicked, the corresponding columns are highlighted in Pipeline Matrix.}

\subsubsection*{Implementation details}

\systemname is implemented as a Python 3 library. The front-end is implemented in Javascript with React~\cite{fedosejev2015react}, D3~\cite{bostock2011d3} and Dagre~\cite{cobarrubia2018dagrejs}. The back-end, responsible for data management, graph merging and the Jupyter Notebook hooks is implemented in Python with Numpy~\cite{walt2011numpy} and NetworkX~\cite{hagberg2008exploring}.

\revision{The \systemname library takes as input a Python array of pipelines in the D3M JSON \cite{D3M2020metalearning, milutinovic2020evaluation} format, and plots the visualization in Jupyter using Widget hooks. The library also enables users to import pipelines from Auto-Sklearn \cite{feurer_auto-sklearn_2019}.} We implemented a bi-directional communication between Jupyter Notebook and our tool. From Jupyter, the user can create an instance of \systemname for their dataset of choice. The main menu (\Figure{fig:teaser}(B)) of \systemname, on the other hand, enables users to subset the data (remove pipelines from the analysis), reorder pipelines according to different metrics, and export the selected pipelines back to Python. The goal of our design is to provide a seamless integration with the existing AutoML ecosystem and Python libraries, and to make it easier for experts to explore, subset, combine and compare results from multiple AutoML systems.

\systemname is already being used in production by the DARPA D3M project members. An open-source release is available at \url{https://github.com/VIDA-NYU/PipelineVis}. 

%% file: text/evaluation.tex
\begin{figure*}[htb]
\centering
\includegraphics[width=0.9\textwidth]{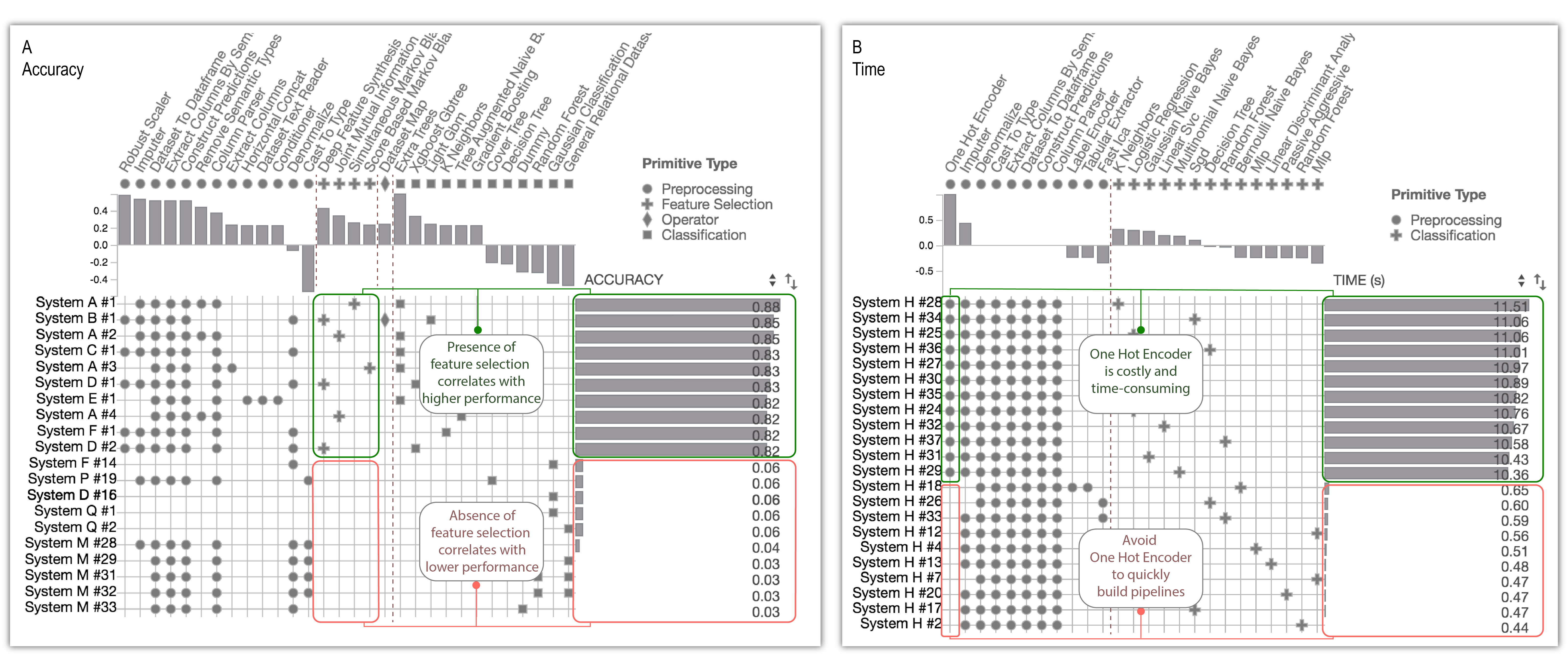}
    \vspace{-.45cm}
\caption{Pipelines Matrix. A) Pipelines are sorted by performance, the  10 best pipelines at the top and the 10 worst pipelines at the bottom. Green  and red boxes show the presence and absence, respectively, of feature selection primitives in the pipelines with their performances. B) Pipelines are sorted by execution time, only pipelines generated by \textit{AlphaD3M} are displayed. Green  and red boxes show the absence and presence, respectively, of one-hot encoder primitive in the pipelines with their execution times.}
    \vspace{-.3cm}
\label{fig:case_study_1}
\end{figure*}

\section{Evaluation} \label{sec:eval}
To demonstrate the usefulness of \systemname, we present case studies that use a collection containing 10,131 pipelines created
as part of the D3M program’s Summer 2019 Evaluation. In this evaluation,  20 AutoML systems were run to solve various ML tasks (classification, regression, forecasting, graph matching, link prediction, object detection, etc.) over 40 datasets, which covered multiple data types (tabular, image, audio, time-series and graph). Each AutoML system was executed for one hour and derived zero or more pipelines for each each dataset. 


\subsection{Case Study 1: Improving an AutoML System}
 To showcase how \systemname supports the effective exploration of AutoML-derived pipelines in a real-world scenario, we describe how an AlphaD3M developer used the system, the insights he obtained, and how these insights helped him improve AlphaD3M. 
 AlphaD3M is an AutoML system based on reinforcement learning that uses a grammar (set of primitives patterns) to reduce the search space of pipelines~\cite{drori_alphad3m:_2018, drori_alphad3m:_2019}. 
 
 The AlphaD3M developer 
 started his exploration using a problem for which 
 AlphaD3M had a poor performance: a multi-class classification task using the \textit{libras move} dataset\footnote{OpenML dataset, https://www.openml.org/d/299} from the OpenML database \cite{OpenML2013}. For this task, in the ranking of all pipelines produced by D3M AutoML systems, the best pipeline produced by AlphaD3M was ranked 18th with an accuracy score of 0.79.

\myparagraph{Comparing pipeline patterns} The developer sought to identify common patterns in the best pipelines that were overlooked by the AlphaD3M search. To this end, he  first sorted the primitives by type and the pipelines by performance. This uncovered useful patterns. 
As \Figure{fig:case_study_1}A shows,  primitives for feature selection were frequently present in the best pipelines, while lower-scoring pipelines did not use these primitives. Although he identified other patterns, the information provided by primitive contribution bar charts indicated that feature selection primitives had a large impact in the score of best pipelines. This information led the developer to hypothesize
that the usage of feature selection primitives might be necessary for pipelines to perform well for the problem and data combination.

\myparagraph{Exploring execution times} %
The developer then analyzed the pipelines produced only by AlphaD3M. \Figure{fig:case_study_1}B clearly shows that pipelines containing one-hot encoding primitives take a substantially longer time
to execute, approximately 10 seconds -- this is in contrast to pipelines that do not use this primitive and take less than 1 second.  
He also saw in primitive contribution bar charts  that the one-hot encoding primitive has the highest impact on the running time. He realized that for this specific dataset, one-hot encoding primitives were used inefficiently, since all the features of the dataset were numeric.
Since an AutoML system needs to evaluate a potentially large number of pipelines during its search, an order-of-magnitude difference in execution time, such as what was observed here,
will greatly limit its ability to find good pipelines given a limited time budget -- for the summer evaluation, this budget was 1 hour.

\begin{figure}[htb]
    \centering
    \includegraphics[width=.6\columnwidth]{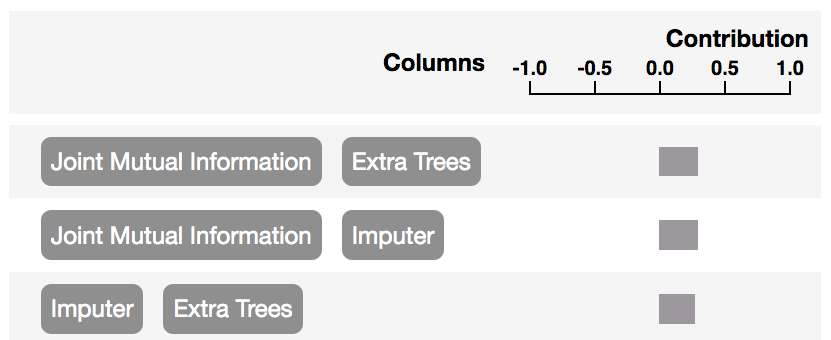}
        \vspace{-.15cm}
    \caption{Combined-Primitive Contribution: the combination of  primitives \textit{Joint Mutual Information}, \textit{Extra Trees} and \textit{Imputer} produce good results together.}
    \label{fig:case_study_1_cpc}
        \vspace{-.5cm}
\end{figure}

\myparagraph{Reducing the search space} AutoML systems have to deal with large search spaces. 
To synthesize pipelines, AlphaD3M takes into account over 300 primitives. This often means that there is a
delay for the system to derive \emph{good} pipelines.
An effective strategy to reduce the search space is the prioritization of primitives. 
In \Figure{fig:case_study_1_cpc}, we can see the results of the Combined-Primitive Contribution view, which shows that the combination of  the primitives \textit{Joint Mutual Information}, \textit{Extra Trees} and \textit{Imputer} produce good results. Using this information, the expert realized that this sequence of primitives could be added to AlphaD3M’s grammar as static components in order to reduce the search space, and consequently, to produce good pipelines faster. 

\myparagraph{Using insights to improve AlphaD3M} After the analysis, the developer modified the AlphaD3M system's handling of feature selection, the use one-hot encoding primitive, and the prioritization of primitives. Feature selection  and  prioritization of primitives were added to the AlphaD3M grammar and  rules were added to the workflow to apply one-hot encoding primitives only for categorical features. The new version of AlphaD3M now leads the ranking  for the multi-class classification task in the  \textit{libras move} dataset with an accuracy of 0.88. \revision{While the AutoML algorithm also has the ability to learn from the
performance of previously derived pipelines, adjustments in the search
heuristic may guide the AutoML to find better solutions faster, or
explore primitive combinations that had not been tried in the past.}
%
With respect to  execution time, the current average time to evaluate each pipeline for this problem is less than 1 second, while previously it took 10 seconds. As a point of comparison, whereas the best pipeline derived  by AlphaD3M after 5 minutes of search had a score of 0.74, now, the best pipeline has a score of 0.79.

\subsection{Case Study 2: Exploring AutoML Approaches}
AutoML systems in the D3M program use different approaches to generate pipelines. In this case
study, we show the use of \systemname to analyze and compare  systems, and discuss some
valuable insights obtained into features that impact a system's performance for a problem.
An AutoML developer set out to compare how six D3M systems -- denoted by \textit{A}, \textit{B},\textit{C},\textit{D}, \textit{E}, \textit{F} -- performed for a regression task
using the \textit{cps 85 wages} dataset.\footnote{OpenML dataset, https://www.openml.org/d/534}
The systems output a total of generated 114 pipelines after 1 hour. 
Since \textit{System F} produced only one pipeline,  it was excluded  from the comparison. 
\textit{System A} obtained the best performance followed by \textit{System B}, \textit{System C}, \textit{System D},  and \textit{System E} with 20.28, 20.29, 20.68, 21.46 and 21.46 mean squared error, respectively.
Using the Pipeline Comparison View, the developer could also easily see noticeable differences in the strategies used by  the AutoML systems  to construct the pipelines. We discuss this further below.

\begin{figure}[t]
    \centering
    \includegraphics[width=\columnwidth]{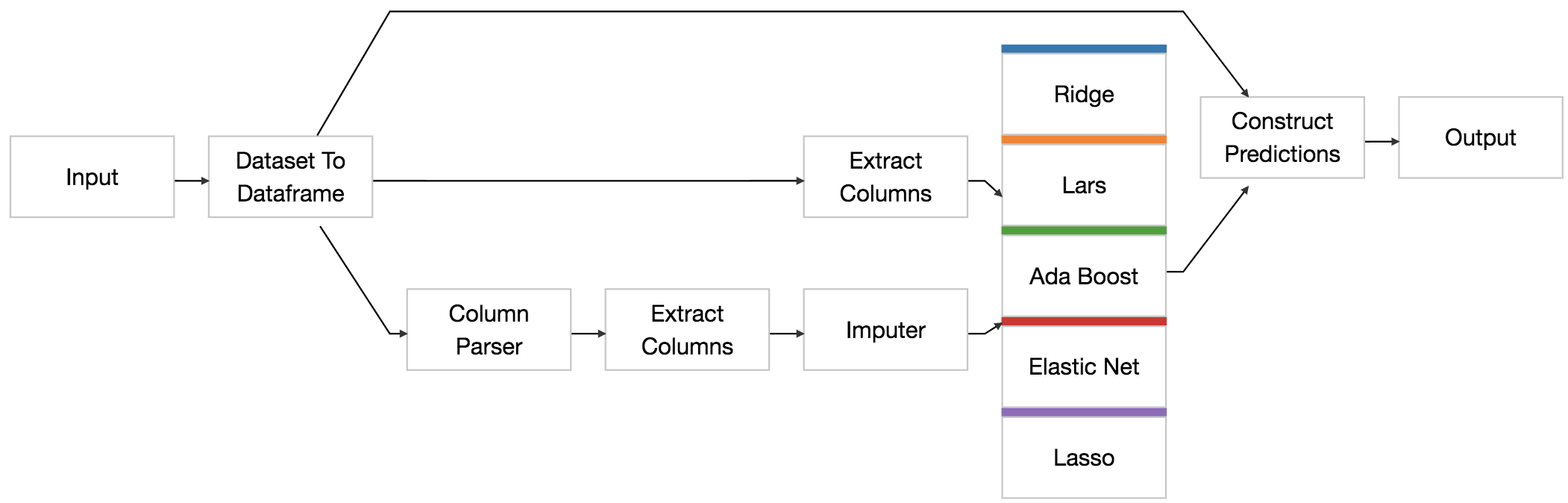} 
    \vspace*{-15pt}
    \caption{A visual comparison of pipelines produced by System D suggests that it fixes the pipeline structure and tries multiple regression algorithms.}
    \label{fig:template_strategy}
\end{figure}

\myparagraph{Template-based approaches} ML templates are  manually designed to reduce the number of invalid pipelines during the search process. Although this approach reduces the search space, it also limits the exploration of potential pipelines.
%
%
\Figure{fig:template_strategy} shows the visual difference for the top-5 pipelines produced by  \textit{System D}.
Note that they all have the same exact structure and only differ in the estimator used (\textit{Ridge}, \textit{Lars}, \textit{Ada Boost}, \textit{Elastic Net} or \textit{Lasso}). A similar behavior was observed for
\textit{System E}.


\begin{figure}[t]
\centering
\includegraphics[width=\columnwidth]{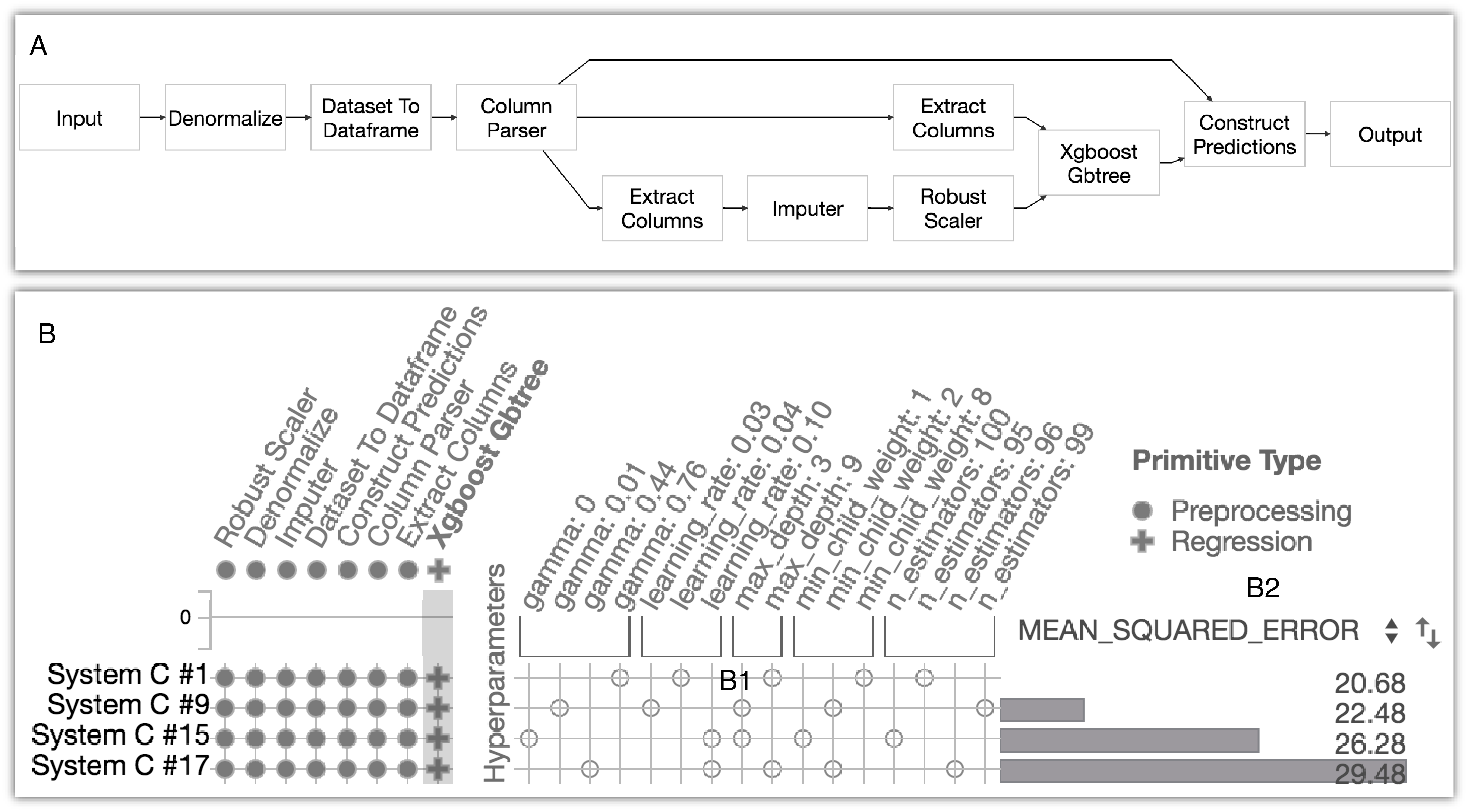}
\vspace*{-15pt}
\caption{\textit{System C} produced four pipelines (\#1, \#9, \#15 and \#17) with the same graph structure, as the merged graph (A) shows. Even though these pipelines have identical structures, the hyperparameter values for the \textit{Xgboost Gbtree} primitive are different (B1), and this results in different scores for the pipelines (B2). This pattern suggests that \textit{System C} tunes the hyperparameter values after it derives the pipeline structure.}
\label{fig:hyperparameter_tuning_strategy}
\end{figure}

\myparagraph{Hyperparameter-tuning strategy} Analyses supported by \systemname can also provide insights into the strategies used by AutoML systems to tune hyperparameters.
While exploring the pipelines produced by \textit{System C}, the developer identified interesting patterns that uncover the strategy this system uses to tune hyperparameters. 
Using the Pipeline Comparison View, he noticed that  some pipelines produced by this AutoML system had the same structure, but used different hyperparameter values. This is illustrated in \Figure{fig:hyperparameter_tuning_strategy}A, which shows the merged graph for 4 distinct pipelines.
He then inspected the hyperparameters of  the \textit{XGBoost} primitive using the one-hot-encoded parameter matrix, and observed that they had different values (see \Figure{fig:hyperparameter_tuning_strategy}B). 
This suggests that \textit{System C} first defines the structure of a pipeline, and then searches for the best-performing hyperparameters values.
We note that the changes in these values have important impact in pipeline performance. For instance, the  mean squared error for the best and worst pipeline are 20.68 and 29.48, respectively.


\begin{figure}[t]
    \centering
    \includegraphics[width=\columnwidth]{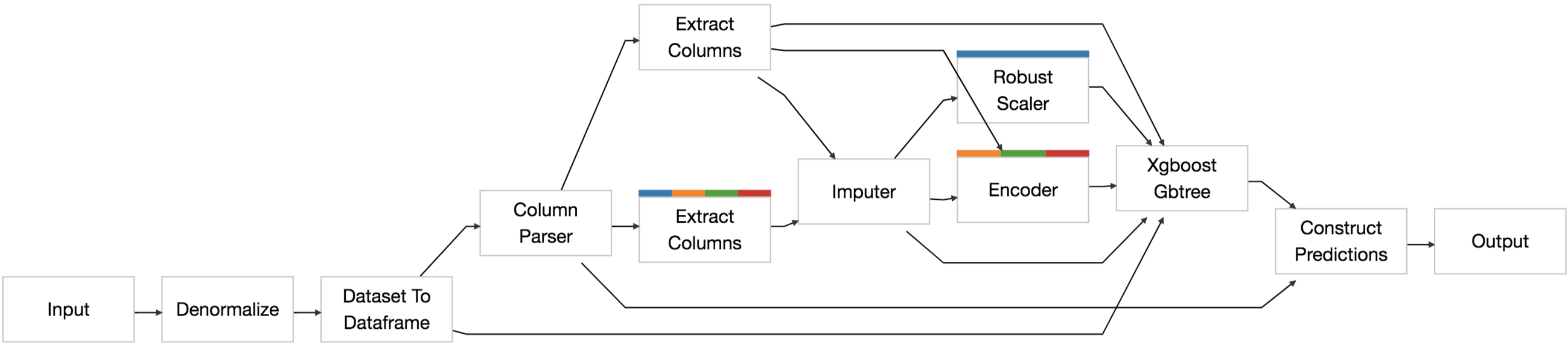}
    \vspace*{-15pt}
    \caption{A comparison of pipelines produced by \textit{System C} indicates that, for a fixed regression algorithm (\texttt{Xgboost}), it searches for alternative sequences of pre-processing primitives.} 
                \vspace{-.25cm}
    \label{fig:pre-processing_search_strategy}
\end{figure}

\myparagraph{Search over pre-processing primitives} 
By exploring another set of pipelines generated by \textit{System C} (see \Figure{fig:pre-processing_search_strategy}), he observed that all pipelines use the same estimator -- the \textit{XGBoost} primitive, but the pre-processing primitives differ -- \textit{Robust Scaler}, \textit{Encoder} and \textit{Extract Columns} are used. 
This suggests that \textit{System C} also searches over alternative pre-processing sequences for a given representative ML estimator, likely in an attempt to optimize the steps for data transformation and normalization. 

\begin{figure}[t]
    \centering
    \includegraphics[width=\columnwidth]{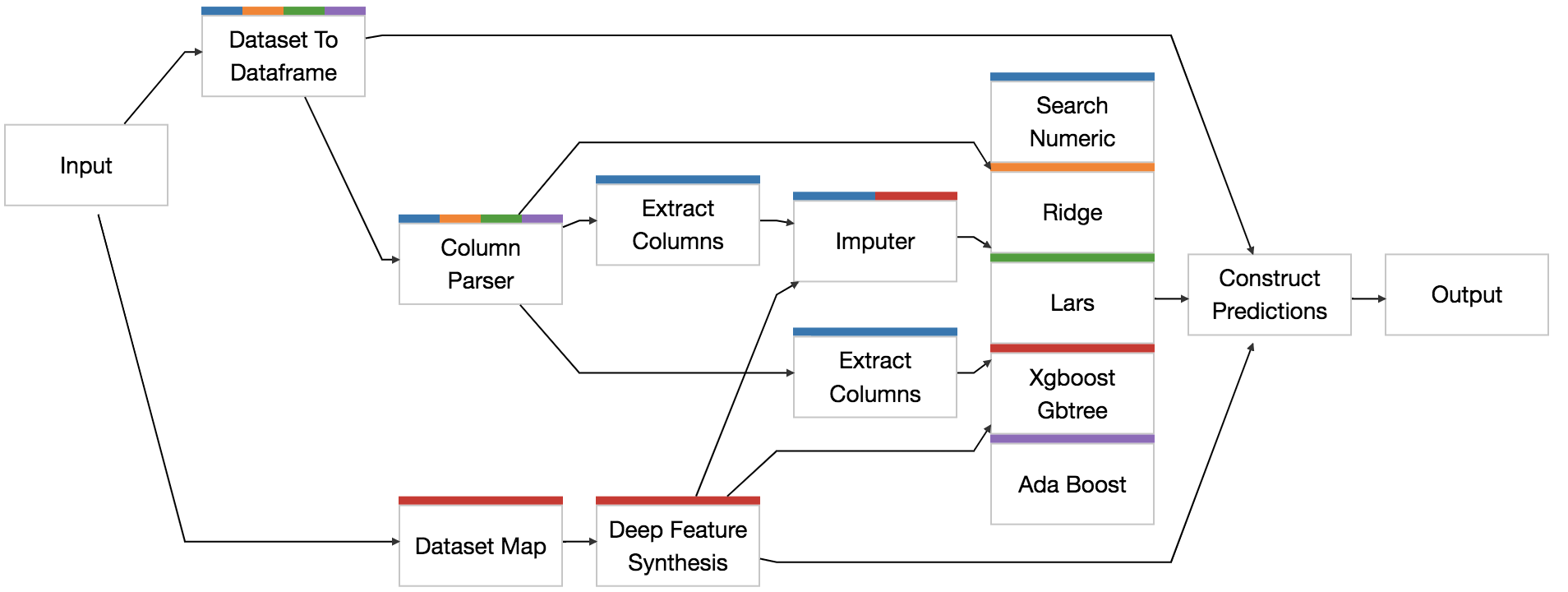}
            \vspace*{-15pt}
    \caption{A comparison of pipelines produced by \texttt{System A} shows that these pipelines vary both in structure and in the primitives used.
    } 
            \vspace{-.45cm}
    \label{fig:full_search_strategy}
\end{figure}

\myparagraph{Full-search approach} The approach applied by \textit{System A} and \textit{System B} seems to search over alternative pre-processing primitives as well as estimators. 
\Figure{fig:full_search_strategy} shows the merged graph for the top-5 pipelines derived by \textit{System A}. Note that these pipelines differ both in the structure and primitives used. Pipelines derived \textit{System B} display a similar behavior.


Although comparing AutoML approaches requires complex analyses, these case studies show that, 
by presenting an overview that highlights differences and similarities for a set of pipelines,
the graph comparison view is quite effective at uncovering interesting patterns that provide insights
into the search strategies employed by AutoML systems.
Additional questions can be explored by drilling down into the details of the hyperparameter values.

The developer also compared the performance of the different systems. 
\textit{System E} and \textit{System D} resulted in the lowest-scoring pipelines, probably due to the unsophisticated search strategies they employ -- the use of fixed templates may be pruning too much of the search space and ignoring efficient pipelines that do not follow the adopted template.
On the other hand, \textit{System A} and \textit{System B}, which perform a broader search, created pipelines
that had high scores.

\subsection{Expert Interviews}

To validate our design decisions, we conducted a second round of interviews with the six  data scientists from the D3M project who had previously helped us identify the system requirements (Section \ref{sec:requirements}).
In the experiment, experts were asked to explore a dataset of their choice according to their usual pipeline exploration workflows. Developers (D1-D4) were asked to use the system to gain insights about the AutoML strategies and identify possible modifications that could result in system improvements. Evaluators (E1, E2) were asked to use the tool to explore the produced pipelines in order to evaluate (and compare) the AutoML systems. They were also asked how this tool could be included in their current workflows to make them more effective.

Each interview took 45 minutes and proceeded as follows. We first presented our system to the participant and clarified any questions they had (10 minutes). Then, we let them choose one problem from the D3M  repository~\cite{D3M2020metalearning} to explore (30 minutes). Finally, we asked if the participant had any  comments on the system (5 minutes). The problems used in this study are shown in Table~\ref{tab:study_datasets}. Note that the systems did not always produce good evaluation metrics for the chosen problems,  indicating that these problems are challening. For example, seven AutoML systems produced 115 pipelines for the \emph{Word Levels} \cite{guzey_classification_2014} classification problem, but no system produced an F-score above 0.33. 

The participants were  free to use \systemname and explore the available pipelines. They were instructed to speak while using the system, following a ``think aloud'' protocol. While the participants performed the task, an investigator took notes related to the actions performed. After completion, the participants filled a questionnaire to express their impressions on the usability of the system. Participants received a \$20 US Dollars gift card for their participation. In this section, we describe the insights gathered by the participants. 

\begin{table*}[t]
\centering
\caption{\label{tab:study_datasets}Datasets used in the expert interviews}
\small
\begin{tabular}{llllllrrl}
\toprule
                                      Dataset & Dataset Type &                Task Type &                   Metric &       Mean Score &       Score Range &  \# Pipelines &  \# Primitives & Participant \\
\midrule
                   Auto MPG \cite{dua2017uci} &      Tabular &               Regression &       Mean Squared Error &  $30.18\pm82.39$ &  $[4.71, 595.24]$ &           103 &             71 &          D1 \\
 Word Levels \cite{guzey_classification_2014} &      Tabular &           Classification &                 F1 Macro &    $0.24\pm0.07$ &    $[0.04, 0.33]$ &           115 &             69 &          D2 \\
             Sunspots \cite{sidc2019sunspots} &  Time Series &  Forecasting &  Root Mean Squared Error &  $27.82\pm23.64$ &  $[8.08, 121.27]$ &           137 &             71 &          D3 \\
   Popular Kids \cite{vanschoren_openml_2014} &      Tabular &           Classification &                 F1 Macro &    $0.39\pm0.06$ &    $[0.21, 0.48]$ &           120 &             64 &          D4 \\
  Chlorine Concentration \cite{chen_ucr_2015} &  Time Series &           Classification &                 F1 Macro &    $0.39\pm0.23$ &    $[0.00, 0.78]$ &            47 &             48 &          E1 \\
 GPS Trajectories \cite{dua2017uci} &      Tabular &           Classification &                 F1 Macro &    $0.48\pm0.15$ &    $[0.00, 0.68]$ &           163 &             91 &          E2 \\
\bottomrule
        \vspace{-.6cm}
\end{tabular}
\end{table*}

\subsubsection*{Expert Insights}

\myparagraph{Data pre-processing} Before they started the investigation, two developers removed outliers from their datasets. D1 and D3 selected datasets with a Mean Squared Error evaluation metric, which is unbounded in the positive real numbers. The two selected datasets, Auto MPG \cite{dua2017uci} and Sunspots \cite{sidc2019sunspots}, had pipelines with error metrics above $100,000.0$, which made the scales difficult to read. In both datasets, the data scientists looked at the Primitive Contribution View and noted that a problem with the \emph{SGD} primitive was likely responsible for these high errors.  They used \systemname subset menu to remove these pipelines from the analysis.

\myparagraph{Performance investigation} Most participants started the analysis by looking at the performance of the pipelines. All developers were interested in how their systems compared against the others. Evaluators, on the other hand, focused on the distribution of scores across all systems. For example, the first comparison E1 did was using the pipeline scores. She grouped pipelines by source and noticed the difference in scores among the top pipelines from each AutoML system. The top two AutoML systems had pipelines with F1 Scores of 0.78 and 0.70, which she mentioned were very close. The other systems produced pipelines with much lower scores, below 0.25.

\myparagraph{Primitive comparison} Participants were very interested in comparing the pipelines produced by different systems. In particular, developers spent a considerable amount of time comparing pipelines from their systems against pipelines from the other  tools. For example, D4 inspected a classification dataset and  found that while a gradient boosting algorithm was used in the top-scoring pipelines, his system was using decision trees. The Primitive Contribution view confirmed his hypothesis that the use of gradient boosting  was indeed  correlated with high scores. He said that he could use this insight to drop and replace primitives in his AutoML search space. D1, D2 and D3 had similar findings in their pipeline investigations.  Evaluators compared primitive usage for a different reason: they wanted to make sure AutoML systems were exploring the search space and the primitives available to their systems. For example, E1 noticed that the best AutoML system used a single classifier type on its pipelines, as opposed to other systems that had more diverse solutions. E2 did a similar analysis on his dataset.

\myparagraph{Hyperparameter search strategy} D1 noticed that the top-five pipelines belonged to the same AutoML system and were nearly identical. She explored the hyperparameters of these pipelines using the  one-hot-encoded hyperparameter view, and found that although they had the same graph structure, they were using different hyperparameters for the \emph{Gradient Boosting} primitive. She compared this strategy with another system which did not tune many hyperparameters, and concluded that tuning parameters was beneficial for this problem.

\myparagraph{Primitive correctness} Participants also used \systemname to check if primitives were being used correctly. A common finding was the unnecessary use of primitives. For example, D2 found that pipelines containing \emph{Cast to Type}  resulted in lower F1 scores. He inspected the hyperparameters of this primitive and noted that string features were being converted to float values (hashes). He concluded that string hashes were bad features for this classification problem, and the \emph{Cast to Type} primitive should be removed from those pipelines. Similar findings were obtained with \emph{One Hot Encoder} used in datasets with no categorical features (D3, E1, E2), and \emph{Imputer} used on datasets with no missing data (D4, E1). E2 also found incorrect hyperparameter settings, such as the use of ``warm-start=true'' in a \emph{Random Forest} primitive.

\myparagraph{Execution time investigation} D4 checked the running times of the pipelines. In particular, he was interested in verifying whether the best pipelines took longer to run. First he sorted the pipelines by score. Then, he switched the displayed metric to ``Time'' and noticed that, contrary to his original hypothesis, the best pipelines were also the fastest. He looked at the Primitive Contribution View in order to find what primitives were responsible for the longer running times, and identified that the \emph{General Relational Dataset} primitive was most likely the culprit. He concluded that if he removed this primitive, he would get a faster pipeline. 

\subsubsection*{Expert Feedback}

We received very positive feedback from the participants. They expressed interest in using \systemname for their work and suggested  new features to improve the system. After the think-aloud experiment, they were asked if they had any additional comments or suggestions. Here are some of their answers: \vspace{-.15cm}
\begin{itemize}
\item D1 mentioned that \systemname is better than her current tools: ``I think this is very useful, we are always trying to improve our pipelines. The pipeline scores can give you some scope, but this is doing it more comprehensively''.
\vspace{-.15cm}
\item D2 liked the debugging capabilities of \systemname: ``Actually, with this tool we can infer what search strategies the AutoML is using. This tool is really nice to do reverse engineering''. 
\vspace{-.15cm}
\item D3 particularly liked the integration with Jupyter Notebook: ``I really liked this tool! It is very informative and easy to use. It works as a standalone tool without any coding, but I can make more specific/advanced queries with just a little bit of code.''
\vspace{-.15cm}
\item D4 wants to integrate \systemname into his development workflow: ``The tools is great, and I as mentioned earlier, it would be even better if an API is provided to ingest the data automatically from our AutoML systems''. E1 and E2 were also interested in integrating this tool with their sequestered datasets, which  used for  evaluation but not shared with the developers
\vspace{-.15cm}
\end{itemize}

\subsection{Usability}

We evaluated the usability of \systemname using the System Usability Score (SUS)~\cite{brooke1996sus}, a valuable and robust tool for assessing the quality of system interfaces \cite{bangor2008empirical}. In order to compute the SUS, we conducted a survey at the end of the second interview: we asked participants to fill out the standard SUS survey, grading each of the 10 statements on a scale from 1 (strongly disagree) to 5 (strongly agree). The SUS grades systems on a scale between 1 and 100 and our system obtained an average score of $82.92 \pm 12.37$. According to Bangor et al.~\cite{bangor2008empirical}, a mean SUS score above 80 is in the fourth quartile and is acceptable. 



%% file: text/conclusion.tex
\section{Conclusions and Future Work} \label{sec:conclusions}

We presented \systemname, a new tool for the exploration of pipeline collections derived by AutoML systems. \systemname advances the state-of-the-art in visual analytics for AutoML in two significant directions: it enables the analysis of pipelines that have complex structure and use a multitude of primitives, and it supports the comparison of multiple AutoML systems.
Users can perform a wide range of analyses which can help them answer common questions that arise when they are debugging or evaluating AutoML systems. Because these analyses are scripted, they can be reproduced and re-used. 

\textbf{Limitations. } \revision{1) Regarding the visualization of hyperparameters, our tool is well suited for categorical data, but numerical parameters suffer from the one-hot-encoded representation. This issue may be addressed by using linked views that show the distribution of numerical hyperparameters, such as in ATMSeer  \cite{wang_atmseer:_2019}. 2) When the pipelines have different structures, it may be hard to see individual graphs on the Pipeline Comparison View. Currently, users need to select individual pipelines to explore them in more detail. As future work, we will investigate interactions that facilitate this process, such as highlighting individual graphs, and groups of nodes/edges.  3) The Combined-Primitive Contribution currently only displays primitive names and their combined effects in the model. More complex analyses can be performed by mining patterns in the data (e.g. \cite{hutter2014efficient, bailis2017macrobase, van2018hyperparameter}).}

\textbf{Future work. }There are many avenues for future work. To increase the adoption of our tool beyond the D3M ecosystem, we plan to add support for other pipeline schemata adopted by widely used AutoML systems.  On the research front, we would like to explore how to \revision{extend the system to support end-users of AutoML systems, who have limited knowledge of ML but need to explore and deploy their models. In particular, we would like to } capture the knowledge derived by users of \systemname and use this knowledge to steer  the search performed by the AutoML system, which in turn, can lead to the generation of more efficient pipelines in a shorter time.
For example, if the user finds that a group  of primitives work well together, they should be able to indicate this to the AutoML system, so that it can focus the search of pipelines that use these primitives.

%% file: text/acknowledgements.tex
\section*{Acknowledgements}

This work was partially supported by the DARPA D3M program and NSF awards CNS-1229185, CCF-1533564, CNS-1544753, CNS-1730396, and CNS-1828576.  Any opinions, findings, and conclusions or recommendations expressed in this material are those of the authors and do not necessarily reflect the views of NSF and DARPA.